\documentclass[12pt]{article}
\usepackage{amsmath} \usepackage{amsfonts} 
\usepackage{natbib,epsfig}
\begin{document}

\makeatletter	   
\renewcommand{\ps@plain}{%
     \renewcommand{\@oddhead}{\textrm{}\hfil\textrm{\thepage}}%
     \renewcommand{\@evenhead}{\@oddhead}%
     \renewcommand{\@oddfoot}{}
     \renewcommand{\@evenfoot}{\@oddfoot}}
\makeatother     

\newtheorem{theorem}{Theorem}
\newtheorem{lemma}[theorem]{Lemma}
\newtheorem{proposition}[theorem]{Proposition}
\newtheorem{corollary}[theorem]{Corollary}

\title{Stochastic adaptation of importance sampler}         
\author{By HENG LIAN}        
\date{\small{Division of Mathematical Sciences, School of Physical \& Mathematical Sciences,\\Nanyang Technological University, \\Singapore 637371\\henglian@ntu.edu.sg}}          
\maketitle

\pagestyle{plain}
\begin{center}
\textbf{SUMMARY }
\end{center}
Improving efficiency of importance sampler is at the center of research on Monte Carlo methods. While adaptive approach is usually difficult within the Markov Chain Monte Carlo framework, the counterpart in importance sampling can be justified and validated easily. We propose an iterative adaptation method for learning the proposal distribution of an importance sampler based on stochastic approximation. The stochastic approximation method can recruit general iterative optimization techniques like the minorization-maximization algorithm. The effectiveness of the approach in optimizing the Kullback divergence between the proposal distribution and the target is demonstrated using several simple examples.

\vspace{0.1in}
\noindent\textit{Some key words}: Adaptive algorithm; Importance sampling; Stochastic approximation.
\section{INTRODUCTION} \label{sec:intr}      
In this paper we are concerned with the approximation of the integral
\[\int h(x)\pi(x)dx=E_\pi h(X)\]
where $\pi$ is a density with respect to the Lebesgue measure on the Euclidean space. When we can sample directly from $\pi$, the simplest Monte Carlo approach for evaluating this integral is to draw independent samples $X_i, i=1,\ldots,N $ from $\pi$ and approximate the integral by the sample mean of $h(X_i)$. When direct sampling is infeasible, the importance sampling approach comes to the rescue by first drawing independent samples from a proposal density $f$, and then use the weighted average 
\[\frac{1}{N}\sum_{i=1}^N h(X_i)w_i\]
to approximate the integral, where $w_i=\frac{\pi(X_i)}{f(X_i)}$. This estimator is unbiased since $E_f[h(X)\frac{\pi(X)}{f(X)}]=E_\pi X$. It is well known that the efficiency of importance sampling depends crucially on the choice of the proposal distribution, since the variance of the estimator is $Var_f(h(X)\pi(X)/f(X))/N$. It is obvious that we can achieve smallest variance with $f\propto |h|\pi$, but this is almost useless in practice. 

The idea of adapting the proposal distribution by utilizing the previously sampled data is a powerful one. While some schemes for adaptation have been devised under the Markov Chain Monte Carlo framework \citep{haario01}, it is generally acknowledged that designing a valid scheme in this situation is a complicated matter and must be carried out carefully in order not to disturb the detailed balance equation. It turns out however that adaptation in importance sampling is much simpler and almost no worries about the validity arise when using changing proposals. This is demonstrated convincingly in \citet{cappe04}. Basically, the change of the proposal can be almost arbitrarily dependent upon previous samples due to the canceling effect on the proposal distribution. Adaptation within the importance sampling framework is generally valid, although genuine effect of adaptation does not always come about. \citet{douc07} showed that a simplistic adaptation scheme within the mixture family of proposal distributions does not achieve the desired effect and the mixture weights stabilize into the uniform weights, while a Rao-Blackwellized version correctly converges to the optimal proposal. Their approach works by drawing $N$ samples from the current proposal, and updating the proposal based on this population. The asymptotic theory is based on the limit theorems when the population size $N$ goes to infinity, while the number of iterations is kept small in practice.

We propose an alternative scheme based on stochastic approximation, by considering a parametric family of proposal distributions. The goal is to update the parameters sequentially to achieve the effect of adaptation. In the next section, we present our approach and state the convergence result based on simplified but stringent assumptions. In section \ref{sec:illu}, we illustrate the method by detailing some concrete circumstances under which our approach works. We also present some simulation results using some simple examples to demonstrate the adaptation ability of our approach. This paper concludes with section \ref{sec:disc}.

\section{STOCHASTIC APPROXIMATION WITH \\WEIGHTED SAMPLES}\label{sec:sais}
Similar to \citet{douc07}, we focus on adaptation with respect to $\pi$ and use the Kullback divergence
\[\int\pi(x)\log\frac{\pi(x)}{f(x)}dx\]
as our criterion for efficiency. Minimizing the divergence is equivalent to maximizing the $\pi$-expected log-likelihood $E_\pi \log f(X)$. For the proposal distributions, we consider a parametric family of densities
\[F=\{f(\cdot|\theta),\theta\in \Theta\subset R^D\}.\]
Our goal is find $\theta$ that makes $E_\pi \log f(X|\theta)$ as large as possible. We assume the maximum is achieved by some $\theta^*\in\Theta$ for simplicity. Maximizing a known function of $\theta$ is usually done by some Newton-like algorithms. We assume such an algorithm exists for maximizing the sampled version of $E_\pi \log f(X|\theta)$, $\sum_i \log f(X_i|\theta)$. Almost all such algorithms can be directly extended to the weighted sample case. Such an algorithm defines a mapping $\theta_{t+1}=M(\theta_t)$, which implicitly depend on the (weighted) samples. We use $M$ to denote both the algorithm and the mapping defined by the algorithm. The notation $M(\theta;\{X_i,w_i\}_1^N)$ is also used to emphasize the dependence of the mapping on the weighted samples.  

For our purpose, suppose we have such an algorithm $M(\theta)$ at hand, which implicitly depends on the weighted samples $(X_1,w_1),\ldots, (X_N,w_N)$. The stochastic approximation importance sampling algorithm is as follows:

\begin{itemize}
\item[-]Start with an initial value $\theta_0$.
\item[-]For $t=0,1,\ldots,T$
\begin{itemize}
\item[-]Draw samples $X_1^t,\ldots, X_N^t$ from $f(\cdot|\theta_t)$.
\item[-]Run the algorithm $M$ on the weighted samples $(X_1^t,w_1^t),\ldots,(X_N^t,w_N^t)$ with $w_i^t=\left(\frac{\pi}{f(\cdot|\theta_t)}\right)(X_i^t)$ to obtain $\tilde{\theta}_{t+1}=M(\theta_t)$.
\item[-]Update $\theta_{t+1}=\theta_t+\gamma_t(\tilde{\theta}_{t+1}-\theta_t)$, where $\{\gamma_t\}$ is a sequence of decreasing positive numbers chosen for the algorithm to converge.
\end{itemize}
\end{itemize}
The convergence of the stochastic approximation algorithm was studied by many authors, including \citet{delyon99, jaakkola94,tsitsiklis94,kushner97}. We choose to work with a maybe overly simplified version here for clarity to avoid any excessive burden in the part of the readers. 

\begin{theorem}(Convergence of stochastic approximation)
Let $\bar{M}(\theta)=\lim_{N\rightarrow\infty}M(\theta;\{X_i,w_i\})$ be the almost sure limit when the number of weighted samples drawn from $f(\cdot|\theta)$ and weighted against the target $\pi$ goes to infinity. Assume 
(1) $\theta_t$ is contained in a convex and compact subset $W$ of $\Theta$;
(2) $l(\theta;x)=\log f(x|\theta)$ is continuously differentiable as a function of $\theta$ and $\partial_\theta E_\pi [l(\theta;X)]=E_\pi[\partial_\theta l(\theta;X)]$;
(3) there exists a unique maximizer $\theta^*$ of $E_\pi [\log f(X|\theta)]$ inside $W$, which is also the unique stationary point;
(4) $\sum_{t=0}^\infty\gamma_t=\infty,\sum_{t=0}^\infty\gamma_t^2<\infty$;
(5) $\langle E_\pi\partial_\theta l(\theta;X), \bar{M}(\theta)-\theta\rangle\le 0,$ for all $\theta\in W$, and it is zero only if $\theta=\theta^*$.

\noindent Then $\theta_t$ converges to $\theta^*$ with probability $1$ as $T$ goes to infinity.
\end{theorem}
The theorem is actually a much simplified version of Theorem 2 in \citet{delyon99}, with $E_\pi[l(\theta;X)]$ acting as the Lyapunov function $V$ in that theorem. The reader can check \citet{delyon99} for the proof and other discussions that greatly relaxed the different assumptions presented here, including the possibility of convergence to other stationary points if they exist. We choose to ignore the complications caused by local maximum or the unfortunate case where the parameters approach the boundary of the parameters space and become unstable. The discussion for the latter concern can also be found in \citet{andrieu05}.

\section{ILLUSTRATION AND SIMULATION}\label{sec:illu}
In this section, we present some concrete instantiations of our general approach described above. We assume all conditions expect condition (5) above are satisfied and will only verify this condition in each case.

\textit{Exponential family }
We consider the one-dimensional exponential family $\{f(x|\mu)=exp\{\eta(\mu)x-\phi(\mu)\}:\mu\in\Theta\}$. Note we choose the mean parameterization so that $E_f(X)=\mu$, and $\theta=\mu$ in this case. The parameter that achieves the minimum divergence is obviously $\theta^*=E_\pi X$. In this simplest case, we do not need to resort to numerical optimization procedure given weighted samples $\{(X_1^t,w_1^t),\ldots,(X_N^t,w_N^t)\}$. We can directly estimate $M(\theta_t)=\frac{1}{N}\sum_{i=1}^N w_i^t X_i^t$, with $\bar{M}(\theta_t)=\theta^*$. That is, the optimal value of $E_\pi l(\theta;X)$ is reached in one single iteration in the ``noiseless" case. With $\bar{M}(\theta_t)-\theta_t=\theta^*-\theta_t$, condition (5) in this case is verified by the concavity of $\log f(x|\theta)$ in $\theta$. The iteration with $N=1$ is 
\begin{equation}\label{eqn:exp}
\theta_{t+1}=\theta_t+\gamma_t\left(\frac{\pi(X_1^t)}{f(X_1^t|\theta_t)}X_1^t-\theta_t\right) ,
\end{equation}
which is just a simple weighted stochastic approximation procedure to find the expected value under $\pi$.

\textit{Stochastic approximation with MM algorithm }
Given a target function $a(\theta):=\sum_i\log f(X_i|\theta)$, MM algorithm is a general technique for iteratively finding the local maximum. In our context, MM stands for minorization-maximization. This algorithm works for weighted version $a(\theta):=\sum_i w_i\log f(X_i|\theta)$ also. Given weighted samples $\{(X_1^t,w_1^t),\ldots,(X_N^t,w_N^t)\}$ with current parameter $\theta_t$, MM algorithm first finds a function $Q(\cdot|\theta_t)$ such that  
\begin{eqnarray*}
a(\theta)=\sum_i w_i^t\log f(X_i^t|\theta)&\ge& Q(\theta|\theta_t), \\
a(\theta_t)=\sum_i w_i^t\log f(X_i^t|\theta_t)&=&Q(\theta_t|\theta_t)
\end{eqnarray*}
The function $Q$ is called the minorizing function of $a$ at the point $\theta_t$.
The new parameter $\theta_{t+1}$ is then chosen to be the maximizer of $Q(\theta|\theta_t)$. The MM algorithm increases the target function monotonically in each iteration since $a(\theta_{t+1})\ge Q(\theta_{t+1}|\theta_t)\ge Q(\theta_t|\theta_t)=a(\theta_t)$. MM algorithm gives us a way to define $M$ given the weighted samples. The corresponding ``noiseless" iteration is defined by $\bar{M}(\theta_t)=\arg\max E_{f(\cdot|\theta_t)}[Q(\theta|\theta_t)]$, where the expectation is taken over the independent weighted samples drawn from $f(\cdot|\theta_t)$. 
To verify condition (5), we make the simplifying assumption that $E_fQ(\theta|\theta_t)$ is a continuously differentiable and strictly concave function. Since $E_fQ(\theta|\theta_t)$ is concave and $\bar{M}(\theta_t)$ is its maximizer, we have $\langle\partial_\theta E_fQ(\theta|\theta_t)|_{\theta=\theta_t}, \bar{M}(\theta_t)-\theta_t\rangle\le 0$. Note that unlike the previous case, here we do not assume $\log f$ itself to be concave, otherwise it is hard to justify the use of MM algorithm. By taking expectations, it is easily verified that $E_{f(\cdot|\theta_t)} Q(\theta|\theta_t)$ is a minorizing function of $E_\pi l(\theta;X)$ at $\theta_t$. Which immediately implies $\partial_\theta E_\pi l(\theta;X)=\partial_\theta Q(\theta|\theta_t)$ at $\theta_t$, and the condition is verified.

\textit{Stochastic approximation with Mixture models }
We consider the family of mixtures models $\{\sum_{d=1}^D \alpha_d p_d(x):\sum \alpha_d=1, \alpha_d> 0\}$, where $p_d(\cdot)$ is fixed and only the mixing weights are adaptable. So $\theta=(\alpha_1,\ldots,\alpha_{D-1})$ in this case. It is well known that the EM algorithm which is popular for mixture models is a special case of MM algorithm. Given the weighted sample $\{(X_1^t,w_1^t),\ldots,(X_N^t,w_N^t)\}$ with current weight parameters $\theta_t$, the new parameter is updated as
\begin{equation}\label{eqn:mix1}
\alpha_d^{t+1}=\sum_i w_i^t \left.\frac{\alpha_d^t p_d(X_i^t)}{\sum_d \alpha_d^t p_d(X_i^t)}\right/N, d=1,2,\ldots,D-1.
\end{equation}
This gives our iterative algorithm $M$, with $[\bar{M}(\theta)]_d=E_\pi \frac{\alpha_dp_d(X)}{\sum_d \alpha_dp_d(X)}$ where $[\theta]_d$ is the $d$-th component of the vector $\theta$. There is an important difference between mixture model inferences using EM and mixture model used as proposal distribution though. For mixture model inferences using EM, generally the component indicator is unknown, and not included in the observations. In importance sampling, the weighted data is something we generate from the current proposal distribution. It is thus possible to take advantage of this and the iterative update can be made using $[M(\theta_t)]_d=\sum_i w_i^t I(Z_i^t=d)/N$ where $Z_i^t$ is the mixture component index of $X_i^t$, which gives the same $\bar{M}$ as before. 

We perform some simple simulations to demonstrate the effectiveness of our adaptation scheme. In all our examples, we choose $\gamma_n\propto 1/n$ although slower decreasing sequence may result in faster convergence rate. We only consider the approximation of $\int\pi(x)dx=1$. In the first example, we let $\pi$ be the standard normal density, and consider proposal distributions from the normal distribution family with adaptable mean and fixed variance of $1$. The number of samples generated from current proposal is $N=1$, and the recursion (\ref{eqn:exp}) is used for stochastic approximation. In the second example, we use the same $\pi$ but the proposal distribution is chosen from the Cauchy family with mean $0$ and adaptable scale $\sigma$. In each iteration we draw two weighted samples (using only one sample will make $\sigma=0$) and use MM algorithm with a quadratic lower bound. The second derivative of $a(\sigma^2)=\sum_{i=1,2}w_i\log f(X_i|\sigma)$ is 
\[\frac{w_1+w_2}{2\sigma^4}-\sum_{i=1,2}w_i\frac{2\sigma^2X_i^2+X_i^4}{(\sigma^2+X_i^2)^2\sigma^4}
\ge -\frac{w_1+w_2}{2\sigma^4}\]
In practice, we assume $\sigma$ is inside a bounded interval so that the second derivative above can be bounded below by a negative constant $C$. Thus $a(\sigma^2)$ is minorized by $a(\sigma_t^2)+(\sigma^2-\sigma_t^2)a'(\sigma_t^2)+\frac{1}{2}(\sigma^2-\sigma_t^2)^2\cdot C$, and the update becomes
\[\sigma^2_{t+1}=\sigma^2_t-\gamma_t C^{-1} a'(\sigma_t^2)\]
This is just a simple stochastic Newton-like update. Note $C^{-1}$ can be absorbed into $\gamma_t$. 

In the third example, we choose $\pi\sim\frac{1}{3}N(-1,1)+\frac{2}{3}N(2,1)$, with the proposal family $\{\alpha N(-1,1)+(1-\alpha)N(2,1)\}$. The iterations start from $\alpha=1/2$ with $N=1$ in each iteration, using the update equation (\ref{eqn:mix1}) where we do not explicitly use the cluster identities.

\begin{figure}
\centering
\makebox{\includegraphics{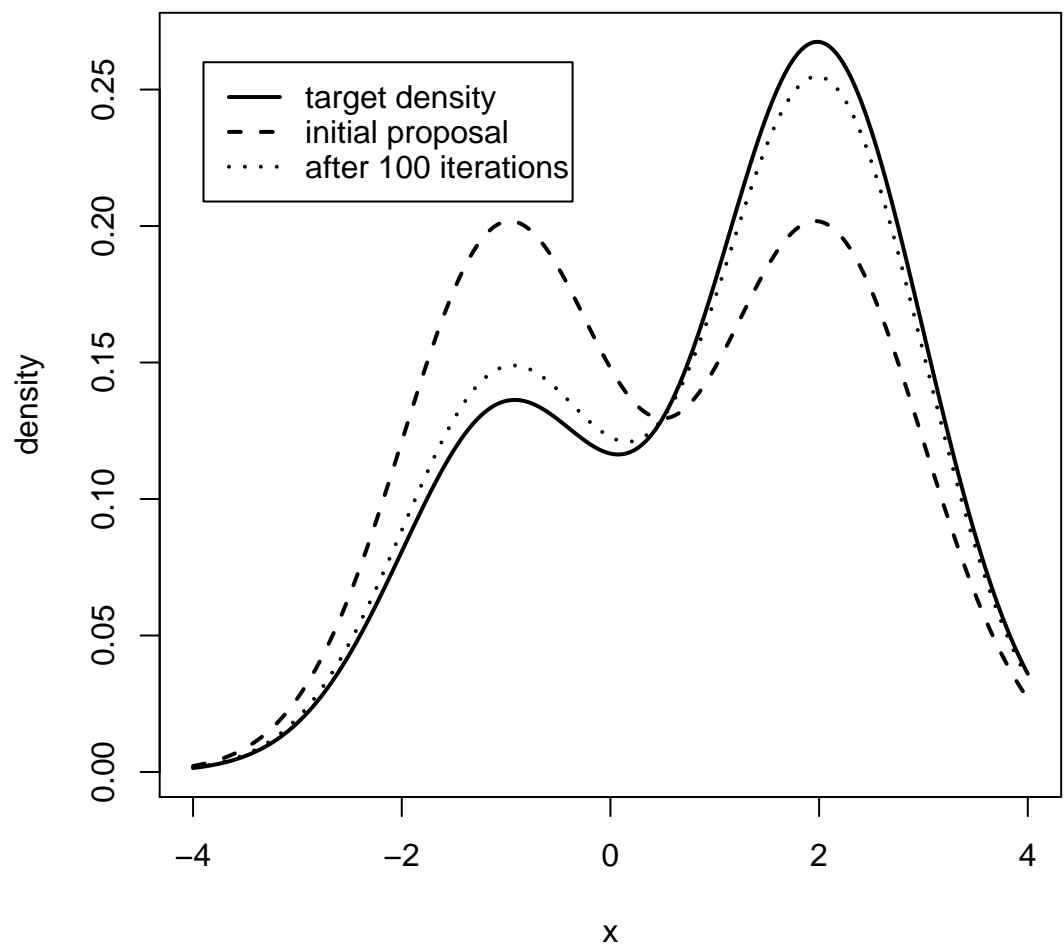}}
\caption{\label{fig:mix}Proposal density after adaptation.}
\end{figure}

For all three examples, we compare the mean square error of adaptive important sampling with that of using a  fixed proposal distribution. The approximation $v_t$ to the integral can be updated using 
\[v_{t+1}=v_t+\gamma_t(\sum_i w_i^t/N-v_t).\] The mean squared errors are reported in table \ref{tab:sim} using 1000 replications with 500 simulated samples (i.e., 500 iterations for the first and the third examples, 250 iterations in the second example). The fixed proposal distribution is chosen to be the same as the initial distribution in the adaptive sampling, which is normal with mean $1$ for the first example, Cauchy with scale $\sigma=2$ in the second example, and mixture of normal with mixing weights $(1/2,1/2)$ in the third example. For each of the three cases, another fixed proposal distribution is chosen to make the mean squared error close to that of adaptive sampling, which is normal with mean $0.1$ in the first case, Cauchy with scale $1.1$ in the second case, and mixture model with mixing weights $(0.35,0.65)$ in the third case. This is denoted by ''fixed proposal 2" in the table. For the mixture family example, we also show in Figure \ref{fig:mix} the density $\pi$ together with the initial proposal density and the proposal density after 100 iterations.  
\begin{table}
\caption{\label{tab:sim}Mean squared errors for the three examples.}
\begin{tabular}{ccccccc}
\\
&&Example 1&& Example 2&&Example 3\\
adaptive sampling            &&0.0000088 &&0.00036 &&0.0000032\\
fixed proposal 1&&0.0003543 &&0.00112 &&0.0000846\\
fixed proposal 2&&0.0000099 &&0.00056 &&0.0000038\\
\end{tabular}
\end{table}

\section{DISCUSSION}\label{sec:disc}
We have proposed an iterative adaption scheme for importance sampler based on stochastic approximation and demonstrated its effectiveness. Our examples only involve independent sampler although kernel-like proposals as used in \citet{douc07} can also be considered. Contrary to the method of \citet{douc07}, where the asymptotics are studied in which the population size $N$ goes to infinity within each iteration, our method works by letting the number of iterations diverge. Other types of asymptotics besides the convergence of the algorithm could be studies as in \citet{delyon99} although this is not our main concern here.

\bibliographystyle{biometrika}
\bibliography{papers,books}

\begin{thebibliography}{8}
\expandafter\ifx\csname natexlab\endcsname\relax\def\natexlab#1{#1}\fi

\bibitem[{Andrieu et~al.(2005)Andrieu, Moulines \& Priouret}]{andrieu05}
\textsc{Andrieu, C.}, \textsc{Moulines, E.} \& \textsc{Priouret, P.} (2005).
\newblock Stability of stochastic approximation under verifiable conditions.
\newblock \textit{Siam Journal on Control and Optimization} \textbf{44},
  283--312.

\bibitem[{Cappe et~al.(2004)Cappe, Guillin, Marin \& Robert}]{cappe04}
\textsc{Cappe, O.}, \textsc{Guillin, A.}, \textsc{Marin, J.~M.} \&
  \textsc{Robert, C.~P.} (2004).
\newblock Population monte carlo.
\newblock \textit{Journal of Computational and Graphical Statistics}
  \textbf{13}, 907--929.

\bibitem[{Delyon et~al.(1999)Delyon, Lavielle \& Moulines}]{delyon99}
\textsc{Delyon, B.}, \textsc{Lavielle, V.} \& \textsc{Moulines, E.} (1999).
\newblock Convergence of a stochastic approximation version of the em
  algorithm.
\newblock \textit{Annals of Statistics} \textbf{27}, 94--128.

\bibitem[{Douc et~al.(2007)Douc, Guillin, Marin \& Robert}]{douc07}
\textsc{Douc, R.}, \textsc{Guillin, A.}, \textsc{Marin, J.~M.} \&
  \textsc{Robert, C.~P.} (2007).
\newblock Convergence of adaptive mixtures of importance sampling schemes.
\newblock \textit{Annals of Statistics} \textbf{35}, 420--448.
\newblock Douc, R. Guillin, A. Marin, J.-M. Robert, C. P.

\bibitem[{Haario et~al.(2001)Haario, Saksman \& Tamminen}]{haario01}
\textsc{Haario, H.}, \textsc{Saksman, E.} \& \textsc{Tamminen, J.} (2001).
\newblock An adaptive metropolis algorithm.
\newblock \textit{Bernoulli} \textbf{7}, 223--242.

\bibitem[{Jaakkola et~al.(1994)Jaakkola, Jordan \& Singh}]{jaakkola94}
\textsc{Jaakkola, T.}, \textsc{Jordan, M.~I.} \& \textsc{Singh, S.~P.} (1994).
\newblock On the convergence of stochastic iterative dynamic-programming
  algorithms.
\newblock \textit{Neural Computation} \textbf{6}, 1185--1201.

\bibitem[{Kushner \& Yin(1997)}]{kushner97}
\textsc{Kushner, H.~J.} \& \textsc{Yin, G.} (1997).
\newblock \textit{Stochastic approximation algorithms and applications}.
\newblock Applications of mathematics ;. New York: Springer.

\bibitem[{Tsitsiklis(1994)}]{tsitsiklis94}
\textsc{Tsitsiklis, J.~N.} (1994).
\newblock Asynchronous stochastic-approximation and q-learning.
\newblock \textit{Machine Learning} \textbf{16}, 185--202.

\end{thebibliography}

\end{document}